# Experimental Demonstration of Dynamic Thermal Regulation using Vanadium Dioxide Thin Films

Ahmed M. Morsy[1], Michael T. Barako[2], Vladan Jankovic[2], Virginia D. Wheeler[3], Mark W. Knight[2], Georgia T. Papadakis[2,4], Luke A. Sweatlock[2], Philip W.C. Hon[2], and Michelle L. Povinelli[1]

*(1) Ming Hsieh Department of Electrical and Computer Engineering, University of Southern California, Los Angeles, USA 90089.*
*(2) NG Next Northrop Grumman Corporation, 1 Space Park Drive, Redondo Beach, California 90278, USA*
*(3) U.S. Naval Research Laboratory, Washington, DC 20375 USA.*
*(4) Department of Electrical Engineering, Ginzton Laboratory, Stanford University, Stanford, California 94305, USA*

## Abstract

We present an experimental demonstration of passive, dynamic thermal regulation in a solid-state system with temperature-dependent thermal emissivity switching. We achieve this effect using a multilayered device, comprised of a vanadium dioxide ($VO_2$) thin film on a silicon substrate with a gold back reflector. We experimentally characterize the optical properties of the $VO_2$ film and use the results to optimize device design. Using a calibrated, transient calorimetry experiment we directly measure the temperature fluctuations arising from a time-varying heat load. Under laboratory conditions, we find that the device regulates temperature better than a constant emissivity sample. We use the experimental results to validate our thermal model, which can be used to predict device performance under the conditions of outer space. In this limit, thermal fluctuations are halved with reference to a constant-emissivity sample.

## 1. Introduction

The use of material design techniques to control the thermal emissive properties of matter has emerged as topic of great interest in current research of intelligent, radiative thermal control. A variety of microstructures have been used for this purpose, including multilayer films [1, 2], microparticles [3], photonic crystals [4], and metamaterials [5, 6].  One particularly interesting application of emissive control is the design of materials that self-regulate their temperature [7, 8], a property we term *thermal homeostasis* [9, 10]. Such a capability is likely to be useful for a variety of applications including satellite thermal control, for which traditional solutions require either electrical power or moving parts [11-14].

The key physical principle required for passive thermal regulation is strong temperature-dependent integrated emissivity. The phase change material vanadium dioxide ($VO_2$), in particular, exhibits a dramatic change to its optical properties across a thermally narrow phase transition [15, 16]. With proper design, $VO_2$-based microstructures can achieve a sharp increase in thermal emissivity across the phase transition temperature near 68 ºC [16, 17]. Intuitively, when the material temperature is below the transition temperature, the emissivity is low, and the object retains heat. When the material temperature exceeds the transition temperature, emissivity increases, and the object loses heat. This negative feedback regulates the material near the temperature of the phase transition [9]. Recent works have demonstrated experimentally broadband emissivity switching for both planar [18] and meta-reflector designs [19]. However, no direct measurement of thermal regulation has been performed. In this paper, we present an experimental method for studying dynamic thermal regulation due to infrared emissive switching. We therefore demonstrate direct evidence of reduction in thermal fluctuations due to emissive switching at the $VO_2$ phase transition.

## 2. Characterization of infrared optical properties of vanadium dioxide

Vanadium dioxide has a phase transition at a critical temperature ($T_c$) of approximately 68 °C [20]. The infrared optical properties of $VO_2$ switch between a low- loss, semi-transparent material (referred to in this article as the insulating state), and a lossy, more reflective material (referred to in this article as the metallic state). Various works in the literature have measured the optical constants of $VO_2$ in the visible and near IR [21-24]. More recent work measured the infrared optical constants of $VO_2$ thin films grown using pulsed layer deposition (PLD) ) [18, 25], sputtering, and sol-gel [26]. It was found that the growth technique influences the optical properties due to the quality of the thin crystalline films [26].

We used atomic layer deposition (ALD) to deposit a $VO_2$ thin film on a Si substrate. Compared to traditional growth methods, ALD allows deposition of highly conformal $VO_2$ films over large areas [27]. The optical constants at temperatures above and below the $VO_2$ phase transition were measured using spectroscopic ellipsometry. The deposition process and measurement method are described in detail in the Methods section, which also lists the ellipsometric fitting parameters. Figure 1 shows the real (*n*, solid lines) and imaginary (*k*, dashed lines) parts of the complex refractive index of the insulating (blue line) and metallic (red line) $VO_2$ states. We observe that both *n* and *k* change significantly between the two states. The higher value of *k* in the metallic state indicates an increase in loss over the entire 2 to 30 μm range.

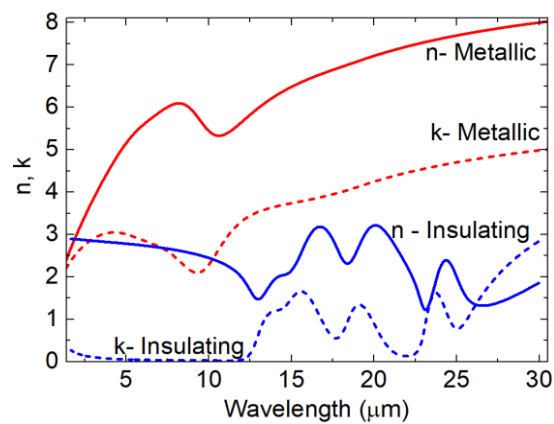

**Figure 1.** Complex dielectric function of metallic (red) and insulating (blue) states of a 98nm-thick vanadium oxide film as determined from spectroscopic ellipsometry measurements. Solid lines give the real part (*n*) and dashed lines give the imaginary part (*k*) of the $VO_2$ refractive index.

## 3. Design of optimized devices for homeostasis

Using the measured optical constants of $VO_2$ shown in Figure 1, we use numerical electromagnetic simulations to optimize our homeostasis device. The figure of merit $\Delta P_{rad}$ is defined as the difference in normalized thermal radiation power between the metallic and insulator states of the device. This quantity is calculated using a full-wave electromagnetic solver, as described in the Methods.

Figure 2 shows that for an isolated $VO_2$ thin film, $\Delta P_{rad}$ is positive for thicknesses less than ~6μm. For experimental convenience, we add a silicon handle layer with a thickness of 200μm (red line in Fig. 2). $\Delta P_{rad}$ is again positive for thickness below ~3μm, with a peak value of 0.22 at a thickness of 800nm. Adding a gold back reflector to the $VO_2$/Si stack enhances the peak value to 0.3 at a smaller $VO_2$ thickness of 75nm. Smaller thicknesses are highly desirable for ALD fabrication.

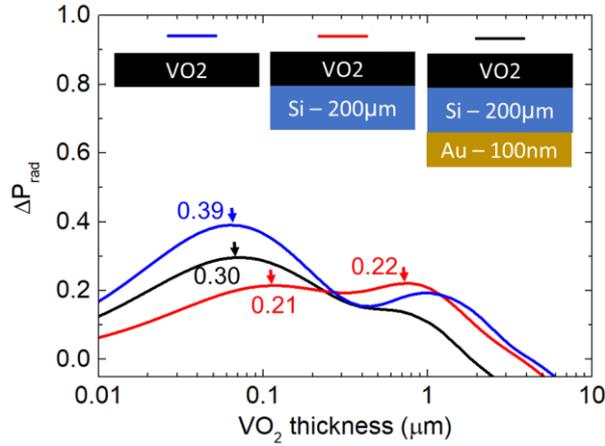

**Figure 2.** Calculated difference in integrated thermal radiation power for an isolated $VO_2$ thin film, $VO_2$ thin film with a handle layer of Si, and $VO_2$/Si/Au multilayer stack. Corresponding schematics and layer thicknesses are shown in the inset.

The use of a gold back reflector also prevents any background thermal radiation from being transmitted through the device, a useful property for thermal homeostasis. Figure 3 shows a comparison between the $VO_2$/Si (Fig. 3(a)) and $VO_2$/Si/Au (Fig. 3(b)) systems. Smoothed lines are superimposed as a guide to the eye. Both structures have higher broadband emissivity in the metallic state of $VO_2$ than in the insulating state (Figs. 3(c) and 3(d)), owing to the increase in optical losses in this state (Fig. 1). A key difference between the two structures, however, is their transmissivity. The $VO_2$/Si/Au structure has zero transmissivity in both the metallic and insulating states. When it is used to cover an external body (e.g. experimental sample holder, or object whose temperature we wish to regulate) the total thermal emission depends only on the emissivity of the $VO_2$/Si/Au stack, not that of the external body. We thus use a gold back reflector in experiments.

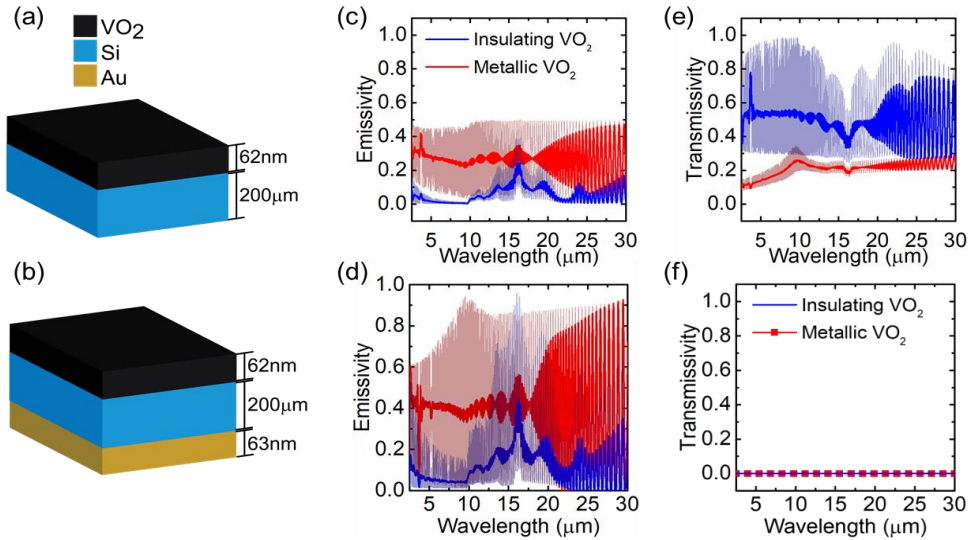

**Figure 3.** (a) Schematic of $VO_2$/ Si, and (b) schematic of $VO_2$/ Si/ Au layered structures with respective simulated (c), (d) emissivity and (e), (f) transmissivity corresponding to $VO_2$ insulating and metallic states.

## 4. Measurement of infrared device properties

We fabricated a $VO_2$/Si/Au device with a $VO_2$ thickness of 62 nm, close to the optimal value calculated in Figure 2. We measured the infrared absorptivity as a function of wavelength for both the insulating and metallic states using FTIR. Figure 4 compares the experimental measurement to smoothed simulation results (see Methods for details). Simulations and FTIR measurements in Figs.

4(a, b) show similar broadband emissivity switching: the emissivity is higher in the metallic state than the insulating state. The integrated difference in radiation power $\Delta P_{rad}$ calculated from the spectra is equal to 0.29 (simulated spectra) and 0.22 (experimental spectra) (Fig. 4(c)). The results suggest that the fabricated sample should emit significantly more heat in the hot (metallic) state, a necessary feature for thermal homeostasis.

Simulations capture most of the measured FTIR spectral features. An offset between the simulated and measured spectra in the insulating state of Fig 4(a) is observed at wavelengths below 15um. This is likely due to a difference between the optical constants of silicon in experiment and the values used in simulation (taken from Ref. [28]).

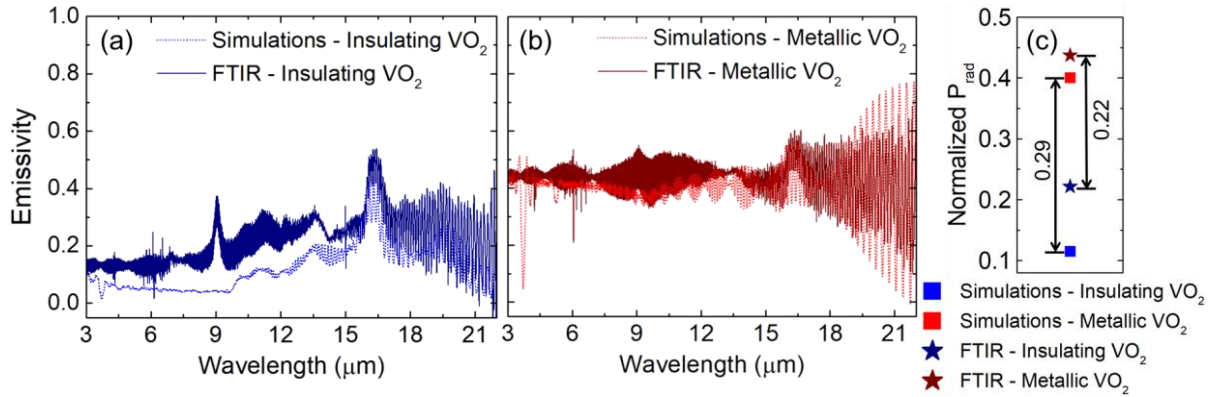

**Figure 4.** Measured and simulated (absorptivity) emissivity spectra of insulating (a) and metallic (b) state of VO2/Si/Au stack. (c) Calculated integrated radiation power normalized to blackbody spectrum.

## 5. Experimental setup and calibration

We designed an experiment to directly test the temperature regulation capabilities of our device. A photograph and a schematic of the experiment are shown in Figures 5(a) and 5(b), respectively. Device samples are mounted on either side of a ceramic heater, containing an embedded thermocouple. The entire structure is suspended in a vacuum chamber, which has an interior black surface to minimize infrared reflection. The chamber is submerged in an ice bath at an ambient temperature of $T_o = 0.5$ °C. The heat load on the sample is varied by changing the input power to the heater, and the resulting temperature is recorded using the thermocouple embedded in the heater.

As shown in the bottom portion of Figure 5(b), the system loses heat through two mechanisms: (1) radiation from the sample, which is the quantity of interest, and (2) parasitic losses that include both radiation from the perimeter of the sample and conduction to the wire leads. To calibrate the parasitic loss, we use gold mirrors with low, constant emissivity ($\varepsilon \approx 0.05$) and measure the temperature rise as a function of applied heat load (yellow circles in Figure 5(c)). The experiment is conducted by using a complete heating and cooling cycle while recording temperature at each steady state. The temperature is first increased in discrete steps, and then decreased again. At each temperature increment, we allow 45 minutes for the system to reach steady state. The temperature-dependent parasitic heat loss function $Q_{loss}(T)$ is determined from a linear fit to this characteristic (see Supporting Information). For each value of applied heat load $Q$, we subtracted the calculated, net radiative loss of gold to obtain $Q_{loss}(T)$, plotted in Fig. 5(d) (see Methods).

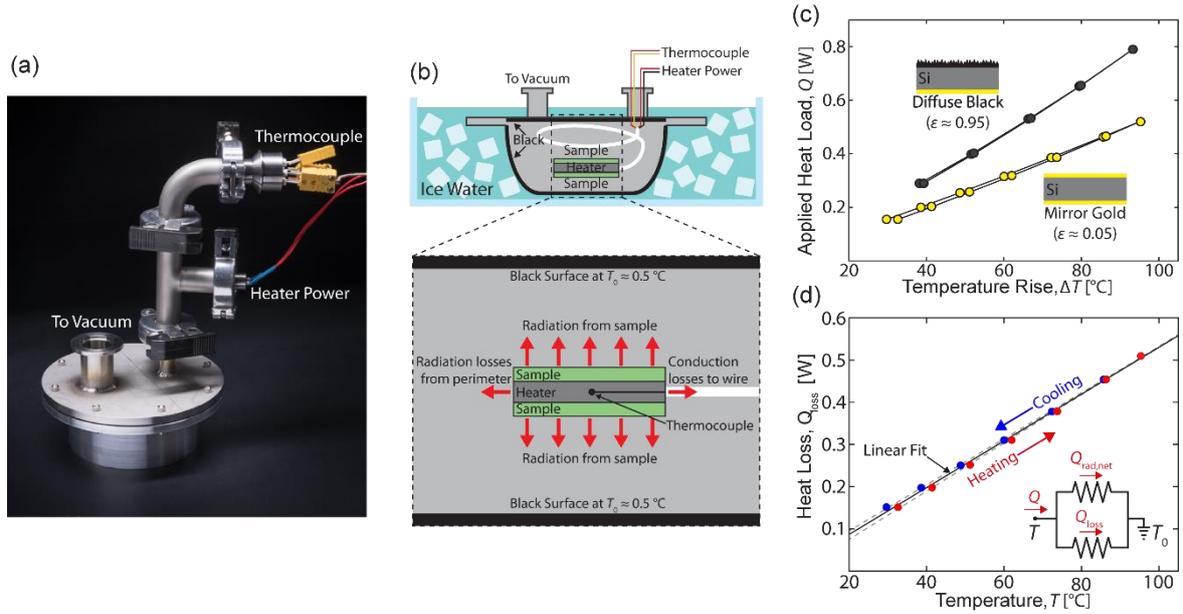

**Figure 5.** Overview of experimental setup. (a) Photograph and (b) schematic of vacuum chamber used to perform thermal measurements. (c) Measured temperature rise as a function of applied heat load for mirror-gold (low emissivity) and diffuse black (high-emissivity) samples, with respect to an ambient temperature of 0.5˚C. (d) Measured parasitic heat loss (sum of radiation losses from perimeter and conduction losses to wire), as a function of temperature, extracted from mirror gold data in (c).

To probe the dynamic range of our measurement system, we also measure a diffuse-black sample with a high total emissivity. Results are shown in Figure 5(c). The data curve for the diffuse-black sample is well separated from the curve for the mirror-gold sample. These two measurements, at the extremes of high and low emissivity, define an operational window for our subsequent, variable-emissivity measurements.

## 6. Measurement of device emissivity

Next, we measure the temperature rise as a function of applied heat load for our VO$_2$ devices. The results are shown in Figure 6(a). The heating and cooling curves trace a hysteresis window around the VO$_2$ phase transition. Inside this window, for constant applied heat load, there is a temperature difference as large as ~5 °C between the heating and cooling curves. The convergence of the heating and cooling curves above and below the hysteresis window suggests that there is negligible temperature drift in the experimental setup. The location of the phase transition can be more readily observed by plotting the derivative of the heat load-temperature curve (Figure 6(a), inset). During heating, the response $dQ/dT$ peaks in the red, shaded region, indicating the transition from insulator to metal at ~80 °C. Upon cooling, $dQ/dT$ peaks at a lower temperature ~60 °C, indicating transition back to the insulating state. We ran this measurement over multiple complete heating/cooling cycles (circles and diamonds) to ensure that there was minimal run-to-run variation in thermal response.

We can use the data from Fig. 6(a) along with the calibration curve in Fig. 5(d) to determine the radiative heat flux emitted by the VO$_2$ sample, $Q''_{rad}(T)$. In steady state, the net heat input to the system is equal to the output:

$$Q + \varepsilon(T)\sigma A T_0^4 = Q''_{rad}(T)A + Q_{loss}(T), \qquad \text{Eq. 1}$$

where the second term on the left-hand side represents the absorptive heat flux at the sample surface due to ambient radiation. This equation can be solved to yield

$$\varepsilon(T) = \frac{Q - Q_{loss}(T)}{\sigma A \left(T^4 - T_0^4\right)},\qquad \text{Eq. 2}$$

assuming that

$$Q''_{rad}(T) = \varepsilon(T)\sigma T^4, \qquad \text{Eq. 3}$$

The radiative heat flux is shown in Figure 6(b). The graph shows that along the heating curve, the radiative heat flux increases sharply near the upper edge of the hysteresis loop. This corresponds to an increase in emissivity. Along the cooling curve, the radiative flux drops at the lower edge of the loop, corresponding to a decrease in emissivity. The inset of Figure 6(b) replots the data to show emissivity as a function of temperature. It can be seen that $\varepsilon_{ins} = 0.22$ in the insulator phase, and $\varepsilon_{met} = 0.46$ in the metallic phase. These values are consistent with those measured using FTIR microscopy ($\varepsilon_{ins} = 0.22$ and $\varepsilon_{met} = 0.44$).

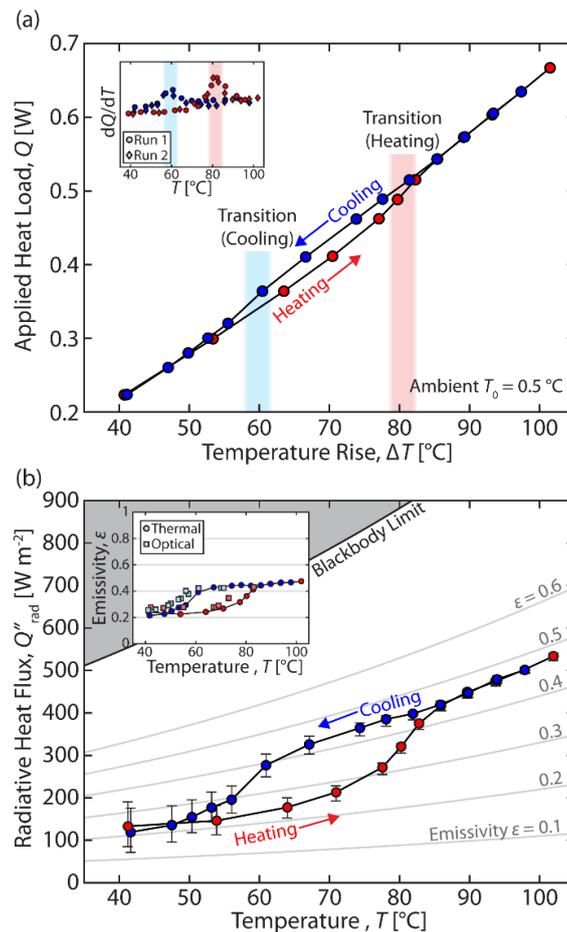

**Figure 6.** (a) Measured temperature rise relative to ambient (horizontal axis) as a function of the applied heat load (vertical axis) for a complete heating and cooling cycle. The inset shows the derivative applied heat load with respect to temperature for two runs (complete heating/cooling cycles). (b) Calculated radiative heat flux from the VO$_2$ surface as a function of temperature. Constant emissivity curves are plotted in grey. The inset shows the effective emissivity of the measured sample as function of temperature.

## 7. Dynamic Thermal Regulation

To demonstrate dynamic thermal regulation, we apply a time-varying heat load and measure the resulting temperature as a function of time. The input power is plotted in Fig. 7(a) and has the form of a square wave with power levels of 0.22 and 0.59W.

For reference, we first measure a near temperature-independent emissivity structure with an alumina top layer. ($Al_2O_3$/ Si/ Au with the corresponding thicknesses of 480nm/ 200μm/ 60nm). The experimental, time-dependent temperature data is shown by the red, dotted line in Fig. 7(b). In response to an increase in input power, the measured temperature rises and then plateaus. When the input power is decreased, the temperature drops again and stabilizes at a lower value. The total range of temperature fluctuation measured is 56°C (red arrows). The measured results can be accurately reproduced using a numerical heat transfer model given by

$$\rho C L_c \frac{dT(t)}{dt} = \frac{Q - Q_{loss}(T)}{A} - \sigma\varepsilon(T)(T^4 - T_0^4), \qquad \text{Eq. 4}$$

where $\rho$ is the effective material density (kg/m³), C is the effective heat capacitance (J/K-kg), $L_C$ is the characteristic length scale of the system (m), and $T_0$= 273.6 K is the ambient temperature. The numerical solution to Eq. 4 is shown by the red, solid line in Fig. 7(b). Physically, the response time of the device is determined by the effective heat capacity, material density, diffusion length and the emissivity of the system. The simulation shows an excellent match to experiment for a fitted value of $\rho C L_C$ =5500 J/(m²-K).

We then measure the performance of our variable-emissivity $VO_2$ device. The experimental data is shown by the blue, dotted line in Figure 7(b). In comparison to the constant-emissivity $Al_2O_3$ device, the total temperature fluctuations are reduced to a value of 50°C. The data can again be well modeled by Equation 4, as shown in Figure 7. Physically, the strong change in emissivity at the phase transition decreases the total temperature fluctuation resulting from a given heat load. This result illustrates the principle of thermal homeostasis.

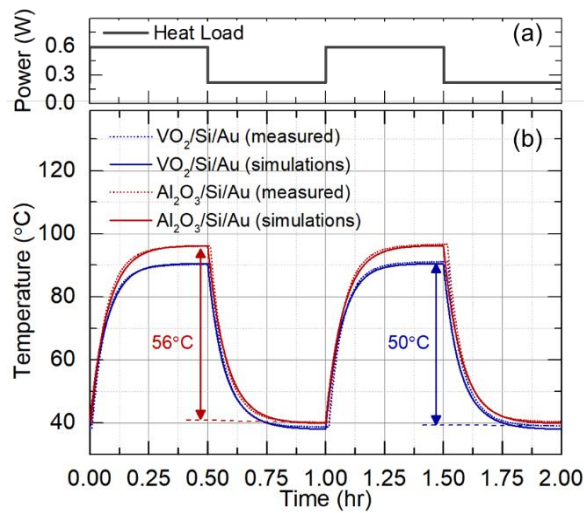

**Figure 7. Thermal Homeostasis.** (a) Square wave time-varying input heat power. (b) measured and calculated response to the input power in for $VO_2$/ Si/ Au (62nm/ 200μm/ 60nm) structure and $Al_2O_3$/ Si/ Au (480nm/ 200μm/ 60nm).

## 8. Discussion

In space applications, under ideal conditions, radiative loss is the only heat dissipation mechanism; parasitic losses vanish. We can use our thermal model to predict the performance of our $VO_2$ device under these conditions. In the absence of parasitic losses, thermal self-regulation of the device is far more effective than under laboratory conditions. We choose input powers of 0.037 W and 0.146 W to ensure that the radiative heat loss from the sample is the same. In this case, the temperature fluctuations in the VO$_2$ device are again around 50°C, as in the experiment of Fig. 7. However, the fluctuations for the constant-emissivity Al$_2$O$_3$ sample are now 108°C. This increase is due to the absence of the parasitic loss pathway. The VO$_2$ sample can therefore self-regulate its own temperature far better than the constant-emissivity sample.

In fact, the magnitude of fluctuations in the VO$_2$ device can be predicted directly from Fig. 6(b). In the absence of parasitic loss, the steady state radiative heat flux is equal to the input power per unit area. From Fig. 6(b), a value of 112 W/m$^2$ corresponds to a temperature of ~39 °C, while a value of 442 W/m$^2$ corresponds to a temperature of ~89 °C. These values correspond well with those obtained in the simulation of Fig. 8(b). For the constant emissivity sample, the temperature fluctuations are much higher. Approximating the Al$_2$O$_3$ sample with a constant emissivity of 0.35, the lower power level corresponds to a temperature of 6 °C, whereas the upper power level corresponds to a temperature of 114 °C, lying well outside the edges of Fig. 6(b). This corresponds to the larger fluctuation of 108 °C seen in Fig. 8(b). A lower bound on the fluctuations is given by the width of the hysteresis curve. For our experimental device, this is close to 20°C. Further improvement in material quality can bring this number down substantially, as observed in literature [29-31]. Another route to performance improvement is to incorporate microstructured designs [6, 9, 10] to increase the total difference in radiated power between metal and insulator states. In this case, for fixed value of temperature fluctuation, the device is expected to accommodate a larger variation in input heat load. The experimental and thermal modeling methods form a general platform for further investigation of dynamic thermal regulation in variable-emissivity systems.

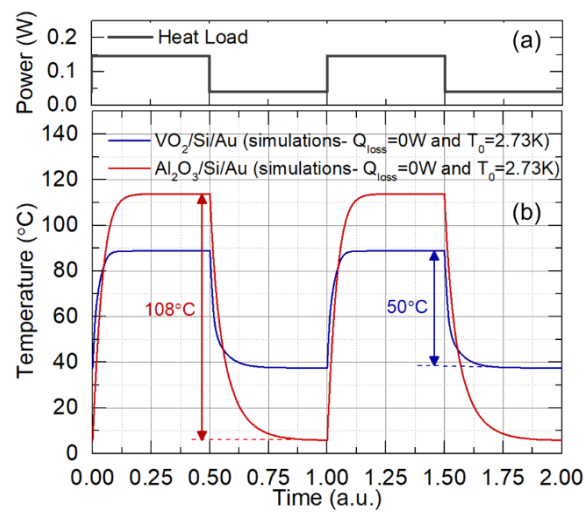

**Figure 8. Thermal homeostasis in space.** (a) Square wave time-varying input heat power. (b) calculated response to the input power in for VO$_2$/ Si/ Au (62nm/ 200μm/ 60nm) structure and Al$_2$O$_3$/ Si/ Au (480nm/ 200μm/ 60nm).

## 9. Conclusion

We have directly demonstrated dynamic, passive thermal regulation via experiments on a VO$_2$ phase-change device. Our device is designed to optimize the increase in radiated power at the phase transition. This trend allows the sample to "self-regulate" its temperature in response to a time-varying, input heat load. Under laboratory conditions, the VO$_2$ device shows a reduction in thermal fluctuations relative to a constant-emissivity device. Using a thermal model, we can extrapolate the device performance to conditions typical of outer space, where radiation is the only heat loss pathway and parasitic losses vanish. Our results demonstrate that emissivity switching can reduce the thermal fluctuations by up to a factor of 2.

Recent investigations [32, 33] have shown flexibility in tuning the phase-transition temperature of VO$_2$ from 28 to 63 °C through doping, addition of dopant atoms, or alloying films. This suggests that various devices could be designed to regulate temperature around fixed values in this range. In terms of ultimate applications, the work presented here provides a key step toward understanding a larger trade space, one that incorporates not only material selection, but also system-level concerns such as payload target temperature and solar heat load.

## 10. Methods

### 10.1 Simulations

The thermal emissivity spectrum $\varepsilon(\lambda, T)$ is calculated using the ISU-TMM package [34, 35], an implementation of the plane-wave-based transfer matrix method. The simulation calculates absorptivity at normal incidence, where absorptivity is equal to emissivity by Kirchoff's law. The wavelength range shown is chosen to be 2–30 μm; outside this range, the blackbody radiance at room temperature is negligible. The normalized thermal radiation power was calculated as

$$P_{rad}(T) = \frac{\int_{2\,\mu m}^{30\,\mu m} d\lambda \cdot I_{BB}(\lambda, T) \cdot \varepsilon(\lambda, T)}{\int_{2\,\mu m}^{30\,\mu m} d\lambda \cdot I_{BB}(\lambda, T)} \qquad \text{Eq. 5}$$

where $I_{BB}(\lambda, T)$ is the blackbody radiance, and $\varepsilon(\lambda, T)$ is the emissivity spectrum.

Matlab Savitzky-Golay filter with an order of 3 and a frame length of 41 was used to smooth the simulated spectra in Figs. 3(c-e) and Fig. 4(a, b). The optical constants for VO$_2$ are taken from the experimental data of Figure 1; the constants for Si and Au are taken from the literature [30].

### 10.2 Fabrication

Amorphous VO$_2$ films (60-120nm) were deposited on 12.5mm x 12.5mm double-side polished, 200 μm thick Si wafers by atomic layer deposition (ALD) in a Veeco Savannah 200 reactor at 150°C using tetrakis(ethylmethyl)amido vanadium and ozone precursors with optimized pulse/purge times of 0.03s/30s and 0.075s/30s, respectively. Under these conditions, the saturated growth rate was 1Å/cycle. All samples of a particular thickness were deposited simultaneously to avoid any run-to-run variation. The thickness was determined using spectroscopic ellipsometry and a general oscillator model previously calibrated with TEM. As-deposited amorphous ALD films underwent an ex-situ anneal at 475°C in 6x10$^{-5}$ Torr of oxygen for 3-4hrs depending on thickness of the film to facilitate the crystallinity required to achieve sharp metal-to-insulator transitions. Raman spectra were collected at room temperature to verify the presence of crystalline, monoclinic VO$_2$ films for all samples after annealing.

### 10.3 Ellipsometer

A VASE JA Woollam spectroscopic ellipsometer was used to characterize the atomic layer deposited VO$_2$ thin films. Ellipsometry measures the complex reflectance ratio of *p* and *s* polarization components, respectively, which may be parametrized by the amplitude component Ψ and the phase difference Δ. Ψ and Δ values for 10 different angles were collected between 55° and 75°. The optical constants were fitted using a series of Lorentzian oscillators in the insulating state with an addition of a Drude oscillator in the metallic state to account for free-electrons in this state, using IR-VASE software.

## 10.4 Fourier Transform Infra-Red (FTIR) Spectroscopy

A Fourier Transform Infrared (FTIR) Spectrometer was used to characterize the reflectance and transmittance of the 62nm VO2/Si/Au multilayer device. We used a Bruker (Hyperion 3000) FTIR attached to Vertex 70 microscope. A 0.5 cm$^{-1}$ resolution and an integration time of 1 second were used. Each measurement was averaged over 5 scans. A ceramic heater (THORLABS™ HT19R) was used to heat the sample, and the temperature was incrementally varied between 25 to 85 °C. For each temperature, the sample was allowed to thermally equilibrate and the interferogram signal was maximized before a measurement was collected.

## 10.5 Thermal Experiment

We use a vacuum chamber with black-painted interior walls submerged in an ice water bath to establish a cold, dark, and low pressure ambient environment (see Figure 5(a-b)). A ceramic resistive heater (Watlow Ultramic, 11.5×11.5×3 mm$^3$, resistance 12 Ω) containing an embedded k-type thermocouple is suspended in the center of the chamber. Two nominally-identical samples (each with a surface area of ~1.65 cm$^2$) are affixed with vacuum grease to either side of the heater to ensure robust thermal contact. The stiff bundle of wires connected to the heater are coiled to suspend the heater in the center of the vacuum chamber. This configuration thermally isolates the heater from the vacuum chamber to minimize parasitic heat losses and promotes isothermal conditions between the heater and the sample. Vacuum is pulled and the chamber is submerged in an ice water bath until interior temperature reaches a stable $T_0 = 0.5$ °C, which is maintained throughout the duration of the experiment.

Once the system is at low vacuum and in thermal equilibrium, we apply incremental changes in heater power and record the steady state temperature at each heat load (see Figure 5c). Each experiment includes a complete heating and cooling cycle that steps up from zero power to maximum power (corresponding to a temperature of 100 °C), and then back down to zero power. This generates a power-temperature characteristic as shown in Fig. 5(c). There is a high sample-to-ambient thermal resistance due to the deliberate thermal isolation of the sample. The high resistance leads to a long thermal time constant, and each data point is collected after 45 minutes when a steady temperature is reached.

The primary parasitic loss in the experiment is due to conduction into the wire bundle that connects the heater to the chamber feedthrough. The rate of heat loss $Q_{loss}$ is independent of the sample being tested and is only a function of the heater temperature $T$. We measure the temperature-dependent heat loss characteristic $Q_{loss}(T)$ for the experimental setup by measuring the relationship between heat dissipation and temperature rise for a set of gold mirror samples with a constant, low emissivity. By letting $Q_{rad,net}$ be defined by the Stefan-Boltzmann equation for a gray body of known emissivity $\varepsilon$ in a vacuum at temperature $T_0$, Equation 1 can be rewritten as

$$Q_{loss}(T) = IV - \sigma \varepsilon A(T^4 - T_0^4) \qquad \text{Eq. 6}$$

where $A$ is the total sample surface area ($A = 3.3$ cm$^2$ in this work), $Q = I \times V$ is the applied Joule heat load, $I$ is the driving current, and $V$ is the voltage drop across the resistive heating element.

To calibrate $Q_{loss}$, we use a low-emissivity sample made using polished silicon with evaporated gold ($\varepsilon \approx 0.05$). We generate the temperature response $T$ as a function of $Q$, as shown in Figure 5(c), across a complete heating and cooling cycle. We then calculate $Q_{loss}(T)$ from Equation 6 and fit the calibration to a linear function, since the range of temperatures is relatively small (less than 100 °C). The calibration curves and extracted loss function are shown in Figure 5(d).

**Acknowledgment**

This work was supported by the National Science Foundation (NSF) grant No. ECCS-1711268. V.D.W. was supported by the Office of Naval Research. The authors thank Bo Shrewsbury for assisting with the thermal measurements.